\newif\ifATCTenPoint
\ATCTenPointfalse

\ifATCTenPoint
  \documentclass[10pt, sigconf, nonacm]{acmart}
\else
  \documentclass[sigconf, nonacm]{acmart}
\fi

\microtypesetup{
  expansion=true,
  protrusion=true,
  tracking=smallcaps,
  kerning=true
}
\usepackage{booktabs}  
\usepackage{caption}   
\usepackage{subfig}    
\usepackage{tikz}      
\usepackage[inkscapepath={C:/Program Files/Inkscape/bin/}]{svg} 

\microtypesetup{protrusion=true, tracking=true} 
\usetikzlibrary{arrows.meta, shapes} 

\title[Ascend-RaBitQ]{Ascend-RaBitQ: Heterogeneous NPU-CPU Acceleration of Billion-Scale Similarity Search with 1-bit Quantization}

\author{Fujun He$^\dagger$, Chuyue Ye$^\dagger$, Huaxiang Cai$^\dagger$, Zetao Lv$^\dagger$, Baolong Cui$^\dagger$,
Wenru~Yan$^\S$, Chao~Zhan$^\S$, Zigang~Zhang$^\S$, Hao Yi$^\dagger$, Jie Xiang$^\dagger$, Xiabing Li$^\dagger$,
Yuhang Gai$^\dagger$, Ziyang Zhang$^\dagger$, Pengfei Zheng$^\dagger$, Yunfei Du$^\dagger$}
\authornote{Corresponding author.}
\affiliation{%
  \institution{$^\dagger$Huawei Technologies Co., Ltd.}
  \institution{$^\S$JD.com}
  \city{}
  \country{}
}

\begin{document}
\pagestyle{plain}  

\begin{abstract}
Vector similarity search is a critical component of modern AI systems, but traditional CPU-based implementations
face fundamental scalability
bottlenecks for billion-scale corpora due to prohibitive computational overhead and memory bandwidth limitations.
While Neural Processing Units (NPUs) offer orders-of-magnitude higher compute density, existing CPU/GPU-optimized 1-bit
RaBitQ quantization implementations cannot be directly ported to NPU architectures due to fundamental hardware mismatches,
and homogeneous design paradigms struggle to simultaneously balance accuracy, memory footprint, and performance.

This paper presents Ascend-RaBitQ, the first heterogeneous NPU-CPU optimized IVF-RaBitQ system for billion-scale vector search,
built on the core insight that decoupling coarse ranking (NPU) from fine ranking (CPU) allows each stage to leverage its optimal hardware,
breaking the long-standing accuracy-memory-performance trade-off. We propose a three-stage heterogeneous execution path comprising
AI Core-accelerated coarse ranking on 1-bit quantized vectors, on-device AI CPU Top-$k$ processing, and host CPU fine re-ranking on full-precision vectors.
We introduce four NPU architecture-native optimizations: fused AIC-AIV operators for parallel distance computation,
computation flow restructuring to exploit rotation orthogonality, fine-grained index block-level load balancing that breaks query boundaries,
and intra-NPU pipeline parallelism between AI Core and AI CPU to mask Top-$k$ latency.
Evaluation on standard datasets shows that Ascend-RaBitQ achieves 3.0$\times$ to 62.8$\times$ faster index construction than the CPU baseline,
up to 11.7$\times$ throughput improvement over the fastest CPU IVF-RaBitQ implementation, and over two orders of magnitude over the mathematically equivalent
CPU baseline, while demonstrating encouraging scalability on distributed multi-NPU systems.
\end{abstract}
\settopmatter{printacmref=false, printccs=false, printfolios=false}

\keywords{Approximate Nearest Neighbor Search,
Billion-scale Vector Retrieval,
1-bit Vector Quantization,
NPU Acceleration,
NPU-CPU Cooperation
}

\maketitle

\section{Introduction}

Vector similarity search has become a critical infrastructure component in modern AI applications, powering retrieval-augmented 
generation (RAG)~\cite{jiang2025rago, shen2025hermes, quinn2025accelerating, kim2026vectorliterag}, search 
and recommendation systems~\cite{reimers2019sentence, barkan2016item2vec, liang2024unify, khattab2020colbert}, and 
computer vision~\cite{fang2022data, frome2013devise, vo2019composing, li2022blip}. 
As large language models (LLMs) and agentic AI systems continue to evolve, the demand for efficient, high-throughput vector 
retrieval has grown exponentially, with industry reports indicating vector databases must now handle billion-scale corpora 
while maintaining millisecond-level latency and cost-effectiveness~\cite{johnson2019billion, shi2025scalable, adams2025distributedann}. 
Traditional vector search systems predominantly run on CPU clusters, employing indexing structures such as inverted file systems (IVF) 
and product quantization (PQ) to reduce search space~\cite{jegou2011product, ge2013optimized, babenko2014inverted}, 
but face fundamental scalability challenges: (1) \textit{computational bottlenecks}---CPU-based distance computations become 
prohibitively expensive as corpus size grows~\cite{johnson2019billion}; (2) \textit{memory bandwidth limitations}---fetching 
high-dimensional vectors from main memory dominates latency, forming a well-documented "memory wall" for vector retrieval 
workloads~\cite{johnson2019billion}.

Neural Processing Units (NPUs) have emerged as transformative accelerators for deep learning workloads, delivering 
orders-of-magnitude improvements in throughput compared to CPUs~\cite{liao2021ascend}. 
This exceptional compute density has driven its widespread adoption in data centers~\cite{huang2025towards, zuo2025serving, deepseekai2026deepseekv4}, 
while making it highly suitable for data-intensive workloads such as vector similarity search.
However, realizing this potential for vector retrieval introduces two fundamental challenges. 
First, NPUs feature limited on-device memory capacity (typically 32-64 GB HBM) compared to CPU clusters with terabyte-scale 
DRAM pools, making it infeasible to store billion-scale high-dimensional vector indices entirely on-device. 
Scaling capacity by adopting multiple NPUs is prohibitively expensive, while simply offloading computation to NPUs 
incurs costly host-device data transfers that erase acceleration gains. Second, while aggressive quantization can reduce 
index size to fit NPU memory, existing schemes such as PQ incur severe accuracy degradation at extreme 
compression ratios, forcing a trade-off between memory footprint and retrieval quality.

Recent advances in quantization theory offer a promising solution: Randomized Bit Quantization (RaBitQ)~\cite{gao2024rabitq} is a 
1-bit quantization scheme with two core characteristics---formal theoretical error guarantees proven to be asymptotically 
optimal, and $32\times$ memory compression compared to \texttt{FP32} vectors. At the same compression ratio, 
RaBitQ achieves substantially higher accuracy than PQ. This extreme compression provides a technically viable 
path to address the NPU memory constraint, reducing billion-scale vector indices from terabytes to tens of gigabytes so they fit 
entirely within a typical Ascend NPU's 64 GB HBM, eliminating costly host-device data transfers. Existing RaBitQ implementations 
have been well optimized for both CPU and GPU platforms: CPU-based designs rely on bitwise operations and SIMD-based batched 
distance estimation~\cite{gao2024rabitq}, while GPU implementations employ GPU-native distance computation schemes, fused CUDA 
search kernels, and GPU-oriented index layouts to achieve over $2.7\times$ higher throughput than IVF-PQ~\cite{shi2026ivfrabitq}. 
However, these CPU/GPU RaBitQ implementations cannot be 
directly adapted to NPU platforms for two critical reasons: their kernels assume execution and memory behaviors not provided by commercial datacenter NPUs, and their homogeneous execution paths do not match the re-ranking required by 1-bit search.

First, existing CPU/GPU RaBitQ implementations are tightly coupled to execution models and memory-system properties that differ 
fundamentally from Ascend NPU. 
CPU-based RaBitQ depends on bitwise operations and SIMD extensions for batched distance estimation. GPU-based RaBitQ is redesigned 
around CUDA's SIMT execution model, coalesced global-memory accesses, hardware-managed caches, and fused GPU kernels. Ascend NPU's Da Vinci architecture adopts a 
fundamentally different design philosophy optimized for AI workloads: it features a heterogeneous compute fabric comprising AI 
Cores (each containing either a Cube Unit for matrix operations or a Vector Unit for element-wise operations) alongside dedicated AI CPUs for general-purpose scalar computation and task scheduling~\cite{liao2021ascend}. 
Unlike GPUs, which support global-memory accesses through a hardware-managed cache hierarchy, Ascend exposes software-managed 
Unified Buffer (UB) and L1 Buffer resources for explicit data movement and on-chip orchestration~\cite{ascend-arch-docs}. 
Together with its static dataflow execution model, this requires algorithm redesign for non-standard vector-search operations 
such as bitwise quantization. Consequently, GPU-native RaBitQ kernels cannot be mechanically ported to Ascend: coalesced memory 
layouts, fused CUDA/SIMT execution, and cache-dependent access patterns must be reformulated around explicit UB/L1 staging, 
AIC/AIV task decomposition, compact-code lookup, and AI Core-AI CPU coordination.

Second, and more fundamentally, homogeneous architectures---whether CPU-only, GPU-only, or NPU-only---fail to balance accuracy, 
memory footprint, and performance simultaneously. Pure CPU implementations struggle to meet throughput requirements for 
billion-scale datasets. Pure GPU implementations face a critical trade-off: 1-bit RaBitQ suffers from lower recall, while 
higher-bit RaBitQ variants~\cite{shi2026ivfrabitq, gao2025practical} quickly exceed limited GPU memory capacity. Similarly, a pure NPU implementation, despite RaBitQ's $32\times$ 
compression fitting the quantized index in HBM, faces an inherent accuracy-performance conflict: 1-bit coarse ranking alone cannot 
deliver production-grade recall without re-ranking, but performing full-precision re-ranking on the NPU would require gathering 
raw vectors from CPU DRAM and transferring them across the host-device boundary---a data movement overhead that erases the NPU's 
acceleration gains. No homogeneous architecture can simultaneously deliver high accuracy, high throughput, and manageable cost 
under practical deployment constraints.

This leads to our core insight: \textit{a heterogeneous co-designed approach}---combining NPU acceleration for fast coarse ranking, 
CPU processing for high-accuracy fine ranking, and joint algorithm-hardware optimization---can break through the long-standing 
accuracy-memory-performance trade-off faced by homogeneous designs. This insight drives the design of \textbf{Ascend-RaBitQ}, 
the first NPU-accelerated RaBitQ library purpose-built for commercial datacenter NPU platforms.

This paper presents the design, implementation, and evaluation of Ascend-RaBitQ. Our key contributions are:

\paragraph{Heterogeneous Three-Stage NPU-CPU Collaborative Execution} We propose a heterogeneous IVF-RaBitQ execution path that 
decomposes query processing across three hardware stages: AI Core for high-throughput 1-bit coarse distance computation, on-device AI CPU for 
low-latency Top-$k$ selection, and host CPU for high-accuracy fine re-ranking on full-precision vectors. This mapping follows the 
dominant bottleneck of each stage: AI Core-AI CPU pipeline parallelism masks Top-$k$ latency, while CPU re-ranking avoids gathering
raw vectors into NPU-contiguous buffers. Controlled ablation experiments show that the implemented optimizations 
deliver a 23.1$\times$ cumulative speedup over the naive baseline. Furthermore, the system achieves up to 11.7$\times$ higher throughput 
than the fastest CPU IVF-RaBitQ implementation, over two orders of magnitude over the mathematically equivalent CPU 1-bit baseline, 
and up to 62.8$\times$ faster index construction.

\paragraph{NPU Architecture-Native RaBitQ Optimizations} We present a RaBitQ implementation optimized for the 
Da Vinci heterogeneous compute architecture. Three key optimizations exploit its unique hardware characteristics: (1) a fused 
AIC-AIV operator that parallelizes matrix multiplication and vector norm calculation across the Cube and Vector units; (2) computation 
flow restructuring that exploits rotation orthogonality to overlap rotation with norm precomputation; (3) adaptation of 1-bit 
distance computation to the CANN native \texttt{gather} operator for FastScan-based computation, which achieves a 15.2$\times$ cumulative speedup 
over the baseline \texttt{select} approach. These 
optimizations collectively unlock the NPU's heterogeneous compute potential for RaBitQ-based vector search.

\paragraph{Efficient Scheduling and Resource Orchestration} We introduce two scheduling techniques tailored for the NPU's 
software-managed memory hierarchy: (1) fine-grained index block-level load balancing that breaks query boundaries, 
achieving a 6.3$\times$ speedup; (2) IVF-aware Top-$k$ pruning that prioritizes closer clusters, 
which combined with intra-NPU AI Core-AI CPU pipeline parallelism masks nearly all Top-$k$ selection latency. Together, they eliminate both computation and memory access bottlenecks in the search 
pipeline.

\paragraph{Production Deployment and Distributed Scaling} We validate Ascend-RaBitQ in real production systems.
We design native distributed scaling for multi-NPU clusters, demonstrating up to 3$\times$ end-to-end throughput 
speedup on 8 NPUs with encouraging distance computation scalability. Ascend-RaBitQ is open-sourced as part of the 
Ascend IndexSDK~\cite{ascend2026indexsdk}.

The remainder of this paper is organized as follows: Section~\ref{sec:preliminary} provides preliminaries on vector search and 
RaBitQ quantization. Section~\ref{sec:methodology} details the NPU IVF-RaBitQ design and implementation. 
Section~\ref{sec:evaluation} presents our experimental evaluation. Section~\ref{sec:related_work} discusses related work. 
Finally, Section~\ref{sec:conclusion} concludes the paper.

\section{Preliminary}
\label{sec:preliminary}

This section provides foundational preliminaries on approximate nearest neighbor search, IVF indexing, the RaBitQ quantization algorithm, and Ascend NPU architecture, establishing the necessary context for our work.

\subsection{Approximate Nearest Neighbor Search and IVF Index}
\label{sec:ann-ivf}

High-dimensional vector similarity search is a core primitive for modern AI systems 
including Agentic AI with industrial deployments now routinely supporting billion- to 
trillion-scale vector corpora~\cite{johnson2019billion,8681160}.

\subsubsection{ANN Problem Definition}
Given a vector corpus $\mathcal{X} = \{\mathbf{x}_1, \mathbf{x}_2, \ldots, \mathbf{x}_N\}$ where each $\mathbf{x}_i \in \mathbb{R}^D$, and a query vector $\mathbf{q} \in \mathbb{R}^D$, the exact nearest neighbor search problem aims to find the vector $\mathbf{x}^{*} \in \mathcal{X}$ that minimizes a distance metric $d(\mathbf{q}, \mathbf{x}^{*})$ (typically Euclidean distance or inner product).
Exact search has $\mathcal{O}(N D)$ complexity per query, which is prohibitively expensive for billion-scale workloads.
Approximate Nearest Neighbor (ANN) search trades minor accuracy loss for orders-of-magnitude performance improvements, and is the standard approach in production deployments.

\subsubsection{IVF Index Structure}
The Inverted File (IVF) index~\cite{jegou2011product} is widely used in billion-scale ANN systems and reduces the search space via vector clustering.

\textbf{Index Construction.}
Run $K$-Means clustering on a training subset of the corpus to generate \textit{nList} IVF centroids,
 each defining an inverted list.
For each base vector $\mathbf{x}_i \in \mathcal{X}$, assign it to the inverted list of its nearest IVF centroid.

\textbf{Query Processing.}
For an input query $\mathbf{q}$, compute its distance to all \textit{nList} IVF centroids, select the top \textit{nProbe} nearest centroids, and search only their inverted lists, typically reducing the search space by one to several orders of magnitude ($nProbe \ll nList$).

IVF is widely adopted in open-source ANN systems including FAISS~\cite{douze2024faiss} and ScaNN~\cite{avq_2020} due to its excellent scalability, low memory overhead, and flexible recall-performance trade-off.

\subsubsection{IVF Variants}
The IVF variants most relevant to this work differ in how vectors are stored in inverted lists:

\textbf{IVF-Flat} stores full-precision raw vectors directly in inverted lists.
For a fixed probing configuration, it avoids vector quantization error and thus achieves the highest recall among these variants, but requires $N \times D \times 4$ bytes for \texttt{FP32} vectors, limiting its applicability to small- to medium-scale datasets.

\textbf{IVF-PQ} combines IVF indexing with PQ~\cite{jegou2011product}, compressing high-dimensional vectors into compact PQ codes (typically 8--32 bytes per vector) and storing the codes in inverted lists.
It reduces memory footprint by 16--32$\times$ compared to IVF-Flat, enabling billion-scale indices to fit in memory, 
but suffers from non-negligible accuracy loss 
due to quantization errors.

\textbf{IVF-RaBitQ.}
The IVF-RaBitQ algorithm adopted in this work retains the memory efficiency advantage of IVF-PQ while significantly reducing 
quantization accuracy loss to address the fundamental trade-off limitation of prior IVF variants.
The RaBitQ quantization method and its integration into the IVF pipeline are detailed in Section~\ref{sec:ivf-rabitq}.

\subsubsection{Alternative Index Structures}
For completeness, we note two alternative index structures:

\textbf{Graph-based indexes} (e.g., HNSW~\cite{malkov2018efficient}) deliver strong recall-latency trade-offs for moderate-scale datasets, but can incur higher memory overhead and challenging billion-scale costs.

\textbf{Disk-based indexes} (e.g., DiskANN~\cite{jayaram2019diskann}) support trillion-scale datasets by offloading indices to SSD, but often incur higher latency than fully in-memory IVF solutions.

\subsection{IVF-RaBitQ Algorithm}
\label{sec:ivf-rabitq}

IVF-RaBitQ combines IVF indexing with RaBitQ 1-bit quantization to enable high-throughput, low-memory vector search.
RaBitQ~\cite{gao2024rabitq} compresses each high-dimensional vector into a compact binary code with formal error guarantees.
We describe the core quantization method and distance approximation scheme.

\subsubsection{RaBitQ Quantization}

For each IVF inverted list with centroid $\mathbf{c}$, RaBitQ first converts each corpus vector $\mathbf{x}_i \in \mathcal{X}$ assigned to that list, and a query vector $\mathbf{q}$ probing that list, into normalized residuals:
\begin{equation}
 \mathbf{u}_{\mathbf{x}_i} =
 \frac{\mathbf{x}_i-\mathbf{c}}{\|\mathbf{x}_i-\mathbf{c}\|_2},
 \quad
 \mathbf{u}_{\mathbf{q}} =
 \frac{\mathbf{q}-\mathbf{c}}{\|\mathbf{q}-\mathbf{c}\|_2}. \label{eq:normalized-residual}
\end{equation}
This normalization separates the scale terms from the angular term used for ranking: the original squared Euclidean distance can be recovered from $\|\mathbf{x}_i-\mathbf{c}\|_2$, $\|\mathbf{q}-\mathbf{c}\|_2$, and the inner product $\langle \mathbf{u}_{\mathbf{x}_i}, \mathbf{u}_{\mathbf{q}}\rangle$.

RaBitQ approximates this normalized inner product with a one-bit code.
Let $\mathbf{P} \in \mathbb{R}^{D \times D}$ be a random orthogonal matrix and $\mathbf{z}_{\mathbf{x}_i} = \mathbf{P}^{\top}\mathbf{u}_{\mathbf{x}_i}$.
The database code stores only the sign of the transformed residual:
\begin{equation}
\mathbf{b}_{\mathbf{x}_i,d} =
\operatorname{sign}(\mathbf{z}_{\mathbf{x}_i,d}) \in \{-1,+1\},
\quad d = 1,\ldots,D. \label{eq:quantize}
\end{equation}
Equivalently, $\mathbf{b}_{\mathbf{x}_i}$ selects the nearest bi-valued codeword in the randomly rotated space.
The index stores this $D$-bit code together with the residual norm $\|\mathbf{x}_i-\mathbf{c}\|_2$ and the database-side normalization constant
\begin{equation}
\alpha_{\mathbf{x}_i} =
\left\langle \frac{\mathbf{P}\mathbf{b}_{\mathbf{x}_i}}{\sqrt{D}}, \mathbf{u}_{\mathbf{x}_i}\right\rangle
= \frac{\|\mathbf{P}^{\top}\mathbf{u}_{\mathbf{x}_i}\|_1}{\sqrt{D}}, \label{eq:rabitq-alpha}
\end{equation}
which are used by the distance estimator during query processing.

\subsubsection{Distance Approximation}

During query processing, for each probed IVF centroid $\mathbf{c}$, RaBitQ transforms the normalized query residual as $\mathbf{z}_{\mathbf{q}}=\mathbf{P}^{\top}\mathbf{u}_{\mathbf{q}}$.
Given the stored database code $\mathbf{b}_{\mathbf{x}_i}$, the normalized inner product is estimated by
\begin{equation}
\hat{s}(\mathbf{q},\mathbf{x}_i) =
\frac{1}{\sqrt{D}\,\alpha_{\mathbf{x}_i}}
\sum_{d=1}^{D} \mathbf{b}_{\mathbf{x}_i,d}\mathbf{z}_{\mathbf{q},d}.
\label{eq:rabitq-ip-estimator}
\end{equation}
This estimator is then substituted into the residual distance decomposition:
\begin{equation}
\hat{d}^{\,2}(\mathbf{q},\mathbf{x}_i) =
\|\mathbf{q}-\mathbf{c}\|_2^2 + \|\mathbf{x}_i-\mathbf{c}\|_2^2
-2\|\mathbf{q}-\mathbf{c}\|_2\|\mathbf{x}_i-\mathbf{c}\|_2
\hat{s}(\mathbf{q},\mathbf{x}_i). \label{eq:approx-dist}
\end{equation}
The query norm term is shared by all candidates in the same probed cluster, while the database residual norm and $\alpha_{\mathbf{x}_i}$ are stored per vector.
In practice, the transformed query can be scalar-quantized, and the summation in Eq.~\eqref{eq:rabitq-ip-estimator} can be implemented using bitwise operations or FastScan-style lookup tables.

\subsection{Ascend NPU Architecture}
\label{sec:npu-arch}
\begin{figure}[t]
\centering
\includegraphics[width=0.95\columnwidth]{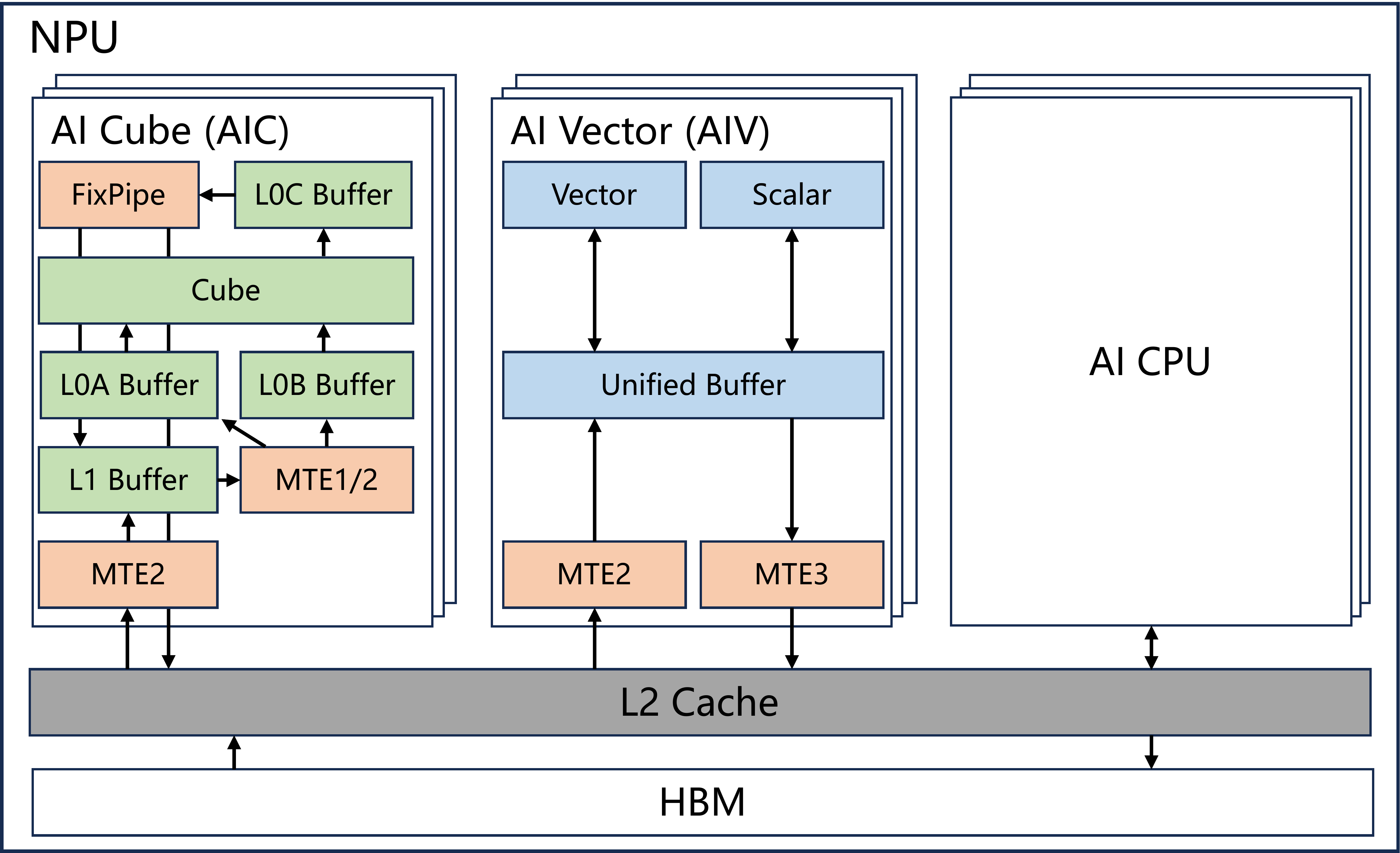}
\caption{NPU Hardware Architecture \cite{ascend-arch-docs}.}
\label{fig:npu_hardware}
\end{figure}

The Ascend NPU family adopts the Da Vinci heterogeneous architecture, designed for high-throughput AI training and inference 
workloads~\cite{liao2021ascend, ascend-arch-docs}. 
Unlike CPUs and GPUs, which rely on hardware-managed caches and relatively uniform compute arrays, the Da Vinci architecture has three defining characteristics that directly shape our system design. An architectural overview is shown in Figure \ref{fig:npu_hardware}.

\subsubsection{Heterogeneous Compute Units at Two Levels}
\label{sec:npu-cube-arch}
The NPU organizes compute resources into two tiers of heterogeneity. 
At the device level, the throughput-oriented AI Core array 
is complemented by on-device AI CPUs---general-purpose scalar processors optimized for latency-sensitive, 
control-heavy tasks such as sorting and scheduling. 
The AI Core array further consists of two physically distinct core types: Cube Unit cores (AICs) for dense matrix 
multiplication and Vector Unit cores (AIVs) for element-wise operations. 
This two-level heterogeneity directly motivates our AIC-AIV fusion operator (Section~3.2.1), 
our computation flow restructuring (Section~3.2.2), 
and our decision to offload Top-$k$ selection to the AI CPU with intra-NPU pipeline parallelism (Section~3.3.3).

\subsubsection{Explicitly Managed On-Chip Buffers} Unlike GPUs, which support global-memory accesses through hardware-managed caches, the Ascend NPU exposes software-managed per-core Unified Buffer (UB) and L1 Buffer resources. Data reaching the compute units must be explicitly loaded through Memory Transfer Engine (MTE) instructions, making performance depend on deliberate data movement and on-chip orchestration. The L2 Cache is hardware-managed and transparent to the programmer, operating analogously to GPU L2 caches. This hybrid memory model is a key reason existing CPU/GPU RaBitQ implementations cannot be directly ported: Ascend-RaBitQ must explicitly stage and organize data in on-chip buffers instead of relying on caches to tolerate irregular access patterns. This motivates our index block-level load balancing and explicit on-chip staging of quantized vectors (Section~3.3.1), as well as our 1-bit vector computation adaptation (Section~3.3.2), which together reduce HBM traffic in the inner loop.

\subsubsection{Dedicated Asynchronous Data Movement Engines} The MTE handles all data movement between HBM and on-chip buffers, operating fully asynchronously with respect to the AI Cores and AI CPUs. Combined with the Huawei Cache Coherence System (HCCS)---a shared-memory interconnect between NPU HBM and host CPU memory---this enables overlapped pipelines spanning both intra-NPU and NPU-CPU boundaries. We exploit this asynchrony to build the intra-NPU pipeline between AI Core and AI CPU (Section~3.3.3), and it provides the hardware foundation for the tighter NPU-CPU collaborative execution path discussed in Section~3.3.4.

In summary, the Da Vinci architecture replaces the uniform compute and automatic caching paradigm of CPUs/GPUs with deliberate heterogeneity and explicit data orchestration. Section \ref{sec:methodology} describes how we translate each of these architectural properties into concrete system optimizations.

\section{Ascend-RaBitQ System Design}
\label{sec:methodology}

\begin{figure*}[htbp]
    \centering
    \includegraphics[width=0.95\textwidth]{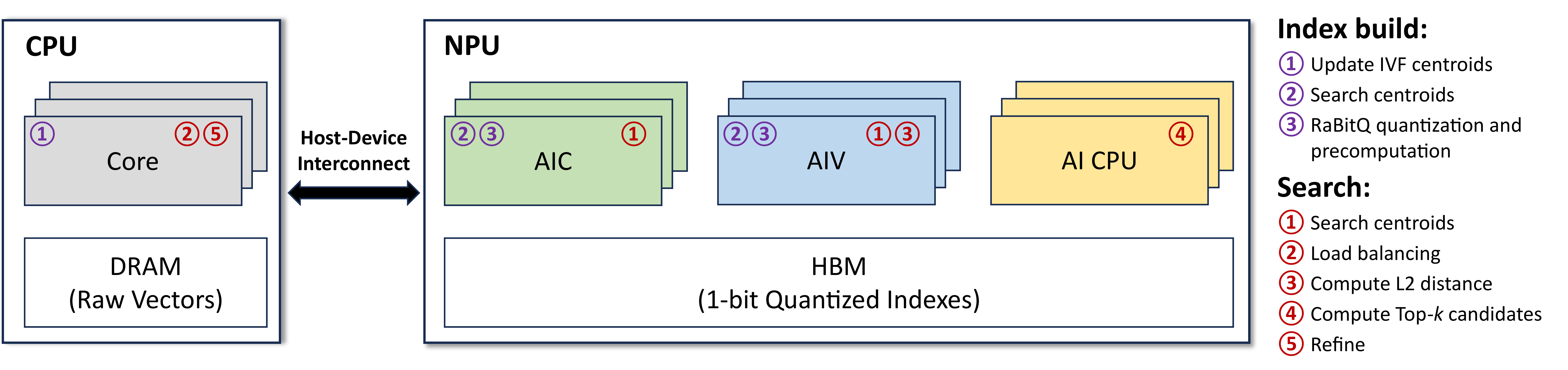}
    \caption{Overview of the Ascend-RaBitQ heterogeneous system architecture and IVF-RaBitQ execution pipeline. The left part shows the CPU-NPU heterogeneous hardware platform, while the right part illustrates how the IVF-RaBitQ algorithm is mapped onto this hardware: NPU handles IVF centroid probing and computation-intensive coarse ranking, while CPU handles global orchestration and final fine re-ranking.}
    \label{fig:ascend_rabitq_overview}
\end{figure*}

Figure \ref{fig:ascend_rabitq_overview} illustrates the system architecture of Ascend-RaBitQ, which implements large-scale, 
efficient IVF-RaBitQ index construction and search based on a heterogeneous NPU-CPU collaborative system. 
Considering the billion-scale vector dataset size and limited HBM capacity, Ascend-RaBitQ adopts a 1-bit quantization design 
instead of multi-bit solutions similar to \cite{shi2026ivfrabitq}. Overall, raw vector data is stored in the high-capacity 
CPU memory system to support billion-scale datasets. The CPU is responsible for global orchestration and re-ranking 
computations on raw vectors to minimize cross-chip data movement. The NPU's HBM stores 1-bit index codes and 
precomputed residual norms and database-side normalization constants, and accelerates computation-intensive vector operations. Ascend-RaBitQ is fully optimized for the 
Ascend NPU hardware architecture, improving index construction and search efficiency through 
three key designs: fused Cube Unit (AIC) and Vector Unit (AIV) collaborative computing, multi-core load balancing scheduling, 
and pipeline latency masking between AI Core and AI CPU. Additionally, Ascend-RaBitQ implements efficient NPU-CPU 
collaboration to boost end-to-end performance.

\subsection{Overall Design of Ascend-RaBitQ}
\subsubsection{Index Construction}
The index construction pipeline consists of three phases: the first two phases iteratively train IVF centroids using $K$-Means and coarse assignment, and the third phase performs core RaBitQ quantization. Details are as follows:
\begin{itemize}
    \item \textbf{Phase 1}: The CPU randomly generates \textit{nList} initial IVF centroids, and iteratively updates these centroids based on the assignment results returned from Phase 2.
    \item \textbf{Phase 2}: The vector dataset is partitioned into multiple batches according to HBM capacity, sequentially offloaded to the NPU for coarse-assignment computation using the latest IVF centroids, and results are transferred back to the CPU. Phases 1 and 2 iterate repeatedly until the $K$-Means termination condition (convergence or maximum iteration limit) is met. The core computation during coarse assignment is batch calculation of L2 distances between dataset vectors and IVF centroids. We design an AIC-AIV fusion operator to improve computational efficiency, which is discussed in detail in Section 3.2.1.
    \item \textbf{Phase 3}: After coarse assignments are finalized, RaBitQ quantization and constant precomputation are performed on all corpus vectors, including normalized residual construction, random orthogonal transformation, 1-bit sign encoding, residual norm calculation, and database-side normalization constant computation. To fully exploit the parallelism between the Cube Unit (AIC) and Vector Unit (AIV), we reorganize the traditional sequential computation flow, which is discussed in detail in Section 3.2.2.
\end{itemize}

\subsubsection{Search Pipeline}
Ascend-RaBitQ processes queries at batch granularity. The first four phases perform coarse search on the NPU based on 1-bit index codes, and the final phase performs re-ranking on the CPU using raw vectors. Details are as follows:
\begin{itemize}
    \item \textbf{Phase 1}: Batch computation obtains \textit{nProbe} nearest IVF centroids for each query vector. This process reuses the coarse-assignment computation flow from index construction Phase~2, fully utilizing both the Cube Unit (AIC) and Vector Unit (AIV).
    \item \textbf{Phase 2}: After the CPU obtains the probed IVF centroids for each query, it performs load balancing scheduling according to batch size, number of \textit{nProbe}, and number of AIV cores to maximize AIV utilization, which is discussed in detail in Section 3.3.1.
    \item \textbf{Phase 3}: Distance calculation is performed, with the core computation being the RaBitQ inner-product estimator between query-side transformed residuals and 1-bit index codes. We compare the CANN \texttt{select}-based direct 1-bit computation path with the \texttt{gather}-based FastScan path, and adopt the latter for higher throughput, with details discussed in Section 3.3.2.
    \item \textbf{Phase 4}: Top-$k$ selection, where we leverage AI CPU computing power for this scalar-dominant computation. Two optimizations are adopted: first, IVF-aware pruning is performed to accelerate sorting; second, pipeline parallelism between AI Core and AI CPU (i.e., Phase~3 and Phase~4) is implemented within the batch to mask Top-$k$ latency, which is discussed in detail in Section 3.3.3.
    \item \textbf{Phase 5}: The CPU performs re-ranking based on raw vectors to select the final Top-$k$ results. We discuss different re-ranking schemes and our NPU-CPU collaboration design in Section 3.3.4.
\end{itemize}

\subsection{Index Construction Acceleration}
\subsubsection{AIC and AIV Fusion Operator}
When calculating L2 distances between a batch of dataset vectors (or query vectors) and a batch of IVF centroids, we decompose the computation into two parallel parts: dot product between dataset vectors and centroids, and norm calculation of each individual vector. The vector dot product is converted into matrix multiplication, where the left and right matrices are dataset vectors and centroids respectively, and we invoke CANN's optimized \texttt{matmul} operator to run on the Cube Unit (AIC). The norm of each vector is calculated on the Vector Unit (AIV) in parallel, and final L2 distance results are aggregated on AIV. We have developed a fused AIC-AIV operator that eliminates synchronization overhead between the two execution units, achieving parallel execution of dot product and norm calculation.

\subsubsection{Computation Flow Restructuring}
In the standard RaBitQ pipeline, each corpus vector is first converted into a normalized residual relative to its assigned IVF centroid, transformed by a random orthogonal matrix, and then encoded by sign extraction to obtain a 1-bit index code. The index also precomputes the residual norm and database-side normalization constant required by the distance estimator. Traditional implementations execute these computations sequentially since they share the same computing unit. Ascend-RaBitQ leverages the NPU's heterogeneous execution units to offload the orthogonal transformation to the Cube Unit (AIC), while decomposing norm and normalization-constant computation to run in parallel where dependencies allow. In particular, we exploit the key property of orthogonal matrices that the transformation preserves vector norms, allowing residual norm calculation to be advanced and overlapped with transformation. In the subsequent stage, similar to Section 3.2.1, dot product computation and remaining element-wise computations are executed in parallel on the Cube Unit (AIC) and Vector Unit (AIV), respectively. This restructured flow fully exploits the parallelism between AIC and AIV.

Ascend-RaBitQ stores the resulting 1-bit index codes in \texttt{UINT8} format, where each bit in a byte corresponds to one dimension of the quantized vector. To match this packed representation and reduce format-conversion overhead, our lookup table (LUT) precomputation also groups 1-bit codes in 8-bit chunks. This differs from the default CPU-oriented RaBitQ scheme~\cite{gao2024rabitq}, which uses 4-bit groups when building lookup tables.

\subsection{Search Acceleration}
\subsubsection{Load Balancing}
\begin{figure}[htbp]
    \centering
    \subfloat[Original query-level kernel scheduling]{\includegraphics[width=0.45\textwidth]{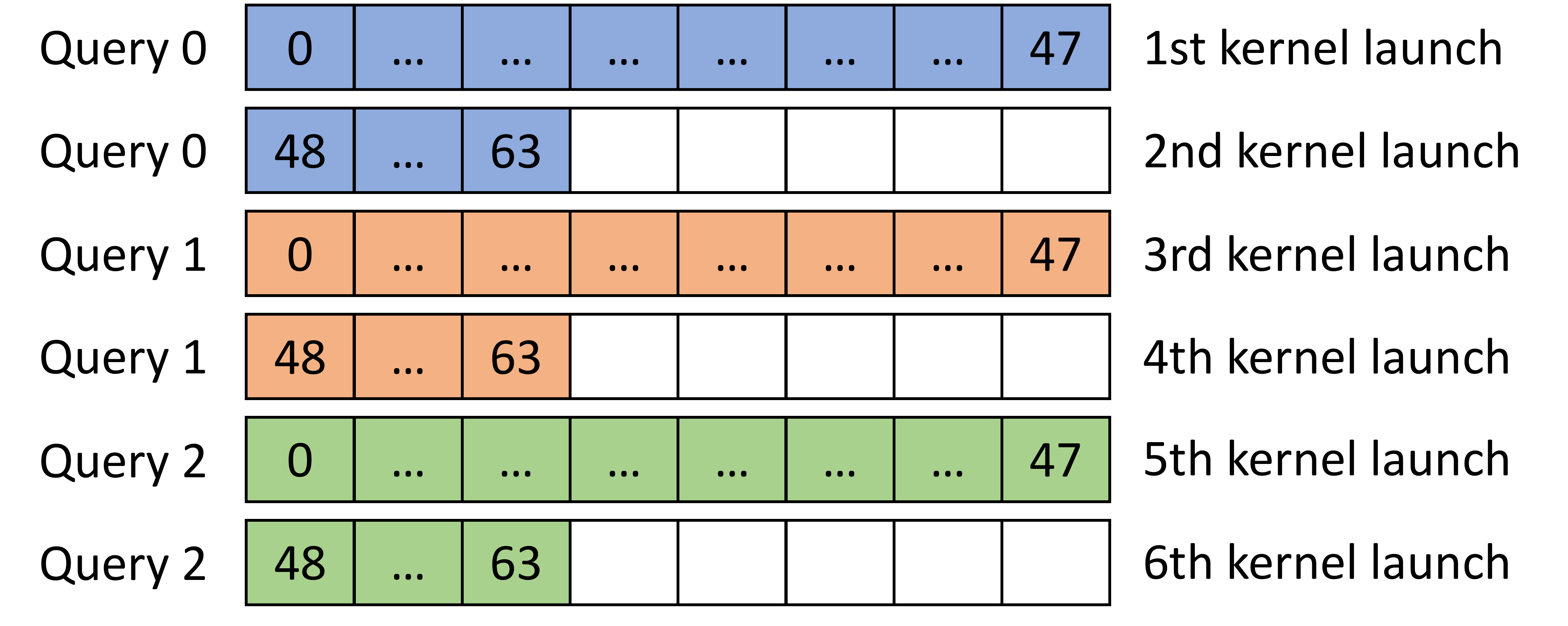}\label{fig:load_balancing_ori}}
    \hspace{0.05\textwidth}
    \subfloat[Optimized index block-level kernel scheduling]{\includegraphics[width=0.48\textwidth]{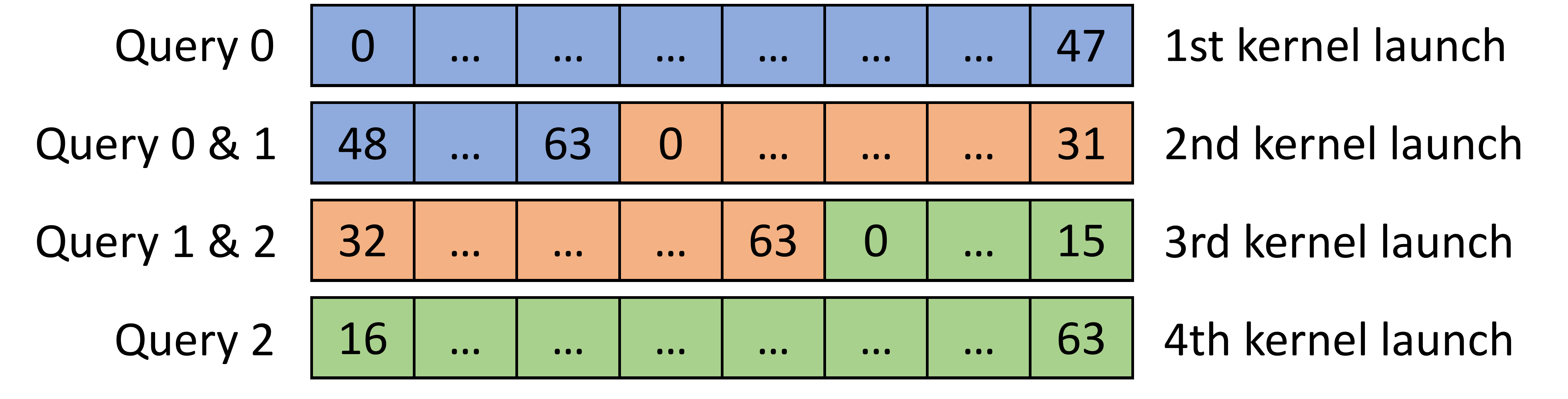}\label{fig:load_balancing_opt}}
    \caption{Load balancing scheduling optimization comparison. The original query-level scheduling causes frequent idle cores across kernel launches for each query, while the optimized index block-level scheduling minimizes idle resources by scheduling tasks at finer granularity across the entire batch.}
    \label{fig:load_balancing}
\end{figure}

Similar to industry IVF solutions such as FAISS~\cite{johnson2019billion}, Ascend-RaBitQ divides each inverted list into \textit{segNum} index blocks. Each index block is stored in contiguous physical addresses internally to improve memory access efficiency, while no contiguity requirement exists between blocks to improve memory utilization. When processing a batch of query vectors, a total of $b \times nProbe \times segNum$ index blocks need to be processed for distance calculation, where $b$ is the query batch size. Existing solutions generally launch distance calculation kernels at query vector granularity: they allocate $nProbe \times segNum$ computation tasks to \textit{coreNum} AIV cores for parallel execution, and launch $\lceil \frac{nProbe \times segNum}{coreNum} \rceil$ kernels sequentially for each query. When $nProbe \times segNum$ is not an integer multiple of \textit{coreNum}, AIV cores will be idle during the last kernel launch for each query, wasting computing power, as illustrated in Figure \ref{fig:load_balancing} (a).

Ascend-RaBitQ breaks the query vector boundary and performs load balancing scheduling at index block granularity. Each kernel launch schedules \textit{coreNum} index blocks and evenly allocates them to AIV cores for computation, as shown in Figure \ref{fig:load_balancing} (b). In this design, AIV computing power waste may only occur during the very last kernel launch for the entire batch, which greatly improves NPU utilization, reduces the total number of kernel launches, and further shortens end-to-end latency.

 We considered an alternative optimization that schedules index blocks from different queries that access the same inverted list into the same kernel launch, reducing index block transfer between the unified buffer and HBM. However, this approach completely disrupts query order, making it impossible to implement pipeline parallelism with subsequent modules as designed in Section 3.3.3 and 3.3.4, ultimately leading to reduced end-to-end performance. Therefore, Ascend-RaBitQ maintains query order when scheduling index blocks.

\subsubsection{1-bit Vector Computation Adaptation}
There are two main categories of distance calculation schemes for RaBitQ: 1) utilizing hardware native 1-bit operation 
capabilities (such as the POPCNT instruction on CPUs) to directly calculate the inner-product estimator between query-side transformed residuals
and 1-bit index codes (CPU implementations require additional scalar quantization of query-side transformed residuals into multiple 1-bit vectors);
 2) the FastScan scheme, which precomputes LUTs and obtains the estimator's partial sums through table 
 lookup~\cite{andre2016cache, andre2017accelerated}. On the Ascend NPU platform, CANN provides both the \texttt{select} operator and the \texttt{gather} operator, corresponding to these two approaches respectively. \texttt{select} takes two sets of vectors as input and uses each dimension of the first vector as a binary mask to select elements from the second vector, realizing the 1-bit direct operation approach. The \texttt{gather} operator implements the FastScan scheme: it first precomputes LUTs and then uses the 1-bit index codes as indices to look up and gather the corresponding distance values from the LUTs.

Although \texttt{select} natively supports \texttt{UINT8} format while \texttt{gather} requires \texttt{UINT8}-to-\texttt{FP32} conversion, our empirical evaluation shows that \texttt{gather}-based FastScan achieves approximately 2$\times$ higher throughput than \texttt{select} (Figure~\ref{fig:ablation_study}). This is because FastScan offloads most computation to the LUT precomputation phase, which is executed once per query and reused across all index blocks. Therefore, Ascend-RaBitQ adopts the \texttt{gather}-based FastScan scheme. We leave a native \texttt{UINT8} \texttt{gather} operator as future work.

\subsubsection{Top-$k$ Optimization}
The Ascend NPU integrates a dedicated AI CPU, which is more suitable for processing scalar-dominant computations compared to the AI Core optimized for parallel vector operations. To fully utilize all NPU computing resources, Ascend-RaBitQ offloads Top-$k$ selection to AI CPU and introduces two targeted optimization techniques. First, when traversing all distance calculation results for sorting, we maintain an upper bound in real time, which is the maximum distance in the current Top-$k$ candidate set. Remaining distances larger than this upper bound are pruned and excluded from further sorting. We exploit the key IVF property that vectors in inverted lists with centroids closer to the query vector are significantly more likely to enter the final Top-$k$ set. Therefore, we traverse inverted lists in the order of the distance between the query vector and their corresponding centroids, prioritizing lists of closer centroids to maximize pruning efficiency.

The second optimization is inter-unit pipeline parallelism between AI Core and AI CPU, 
as illustrated in Figure \ref{fig:pipeline_overview} (a) and (b). 
Specifically, within a batch, when a query vector completes distance calculation on AI Core, a dedicated flag in HBM is atomically set. The AI CPU continuously scans these flags and immediately starts Top-$k$ computation for a query as soon as its corresponding flag is set. Our evaluation shows that this pipeline parallelism effectively masks nearly all Top-$k$ latency, shortening end-to-end search latency significantly.

\subsubsection{NPU-CPU Collaboration}
\begin{figure}[htbp]
    \centering
    \includegraphics[width=0.85\linewidth]{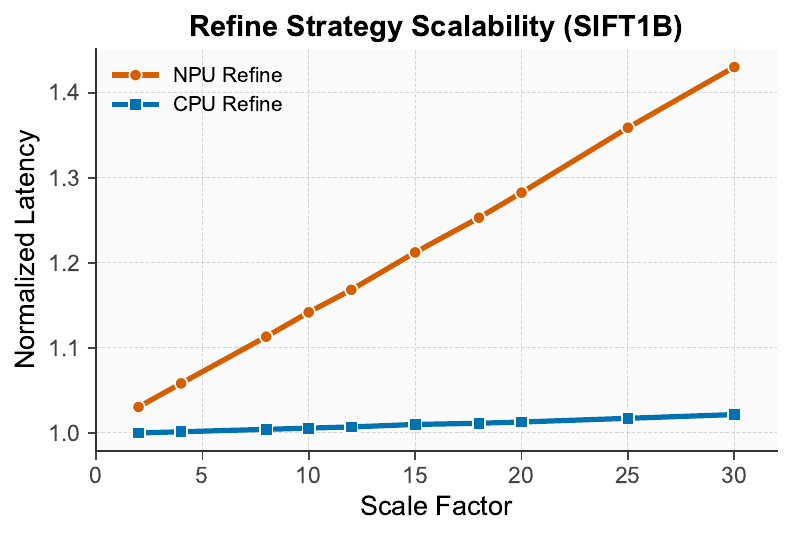}
    \caption{Comparison of NPU re-ranking vs. CPU re-ranking end-to-end latency across different scale factors on SIFT1B~\cite{johnson2019billion} dataset. CPU re-ranking latency remains stable as scale factor increases, while NPU re-ranking latency grows significantly due to CPU data gathering overhead from discontinuous memory.}
    \label{fig:refine_scalability}
\end{figure}

Ascend-RaBitQ adopts extreme 1-bit quantization to fully utilize limited HBM capacity, requiring a final re-ranking step on raw vectors to achieve acceptable recall rate. One candidate re-ranking approach is to send required raw vectors from CPU to NPU, and perform re-ranking on NPU to leverage its high vector computing throughput. The alternative is to perform re-ranking on CPU directly, avoiding cross-device data transfer. Figure~\ref{fig:refine_scalability} compares the end-to-end latency of these two re-ranking strategies across different scale factors of Top-$k$. It shows that CPU re-ranking maintains stable latency as the scale factor increases, while NPU re-ranking latency grows sharply and far exceeds the CPU approach. The primary bottleneck lies in the CPU data preparation step: to send the data, the CPU first needs to gather raw vectors from discontinuous memory space into contiguous buffers according to the returned Top-$k$ coarse results, which becomes a significant performance bottleneck, especially when the Top-$k$ value increases, as shown in Figure~\ref{fig:refine_scalability}. Therefore, Ascend-RaBitQ performs re-ranking on CPU to eliminate unnecessary data movement.

Although the computation amount of re-ranking is reduced by several orders of magnitude compared to coarse ranking, 
considering the limited computing power of CPU and the possibility that CPU may simultaneously process other 
vector database control tasks, CPU re-ranking may still become an end-to-end bottleneck under high load. 
Currently, Ascend-RaBitQ adopts a sequential execution model: the CPU waits for the NPU to complete all coarse 
ranking and Top-$k$ selection before starting re-ranking. We are engineering a tighter NPU-CPU collaborative 
execution path to mask re-ranking latency under high load. In this design, the CPU monitors completion flags 
stored in HBM and immediately starts re-ranking for a query as soon as its coarse results are available. 
Shared-memory signaling between NPU and CPU, natively supported in Ascend and Kunpeng servers via the 
Huawei Cache Coherence System (HCCS) interconnect, provides one efficient implementation path for this design.

\begin{figure}[htbp]
    \centering
    \subfloat[No pipeline parallelism across heterogeneous hardware]{
        \includegraphics[width=0.95\columnwidth, height=3.5cm, keepaspectratio]{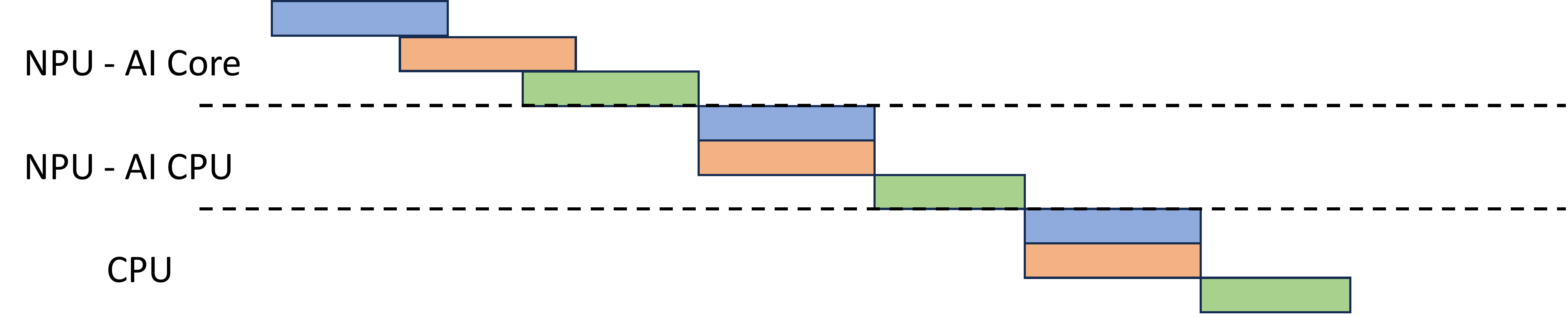}
        \label{fig:pipeline_no}
    }\\[5pt]
    \subfloat[Pipeline parallelism only within NPU between AI Core and AI CPU]{
        \includegraphics[width=0.95\columnwidth, height=3.5cm, keepaspectratio]{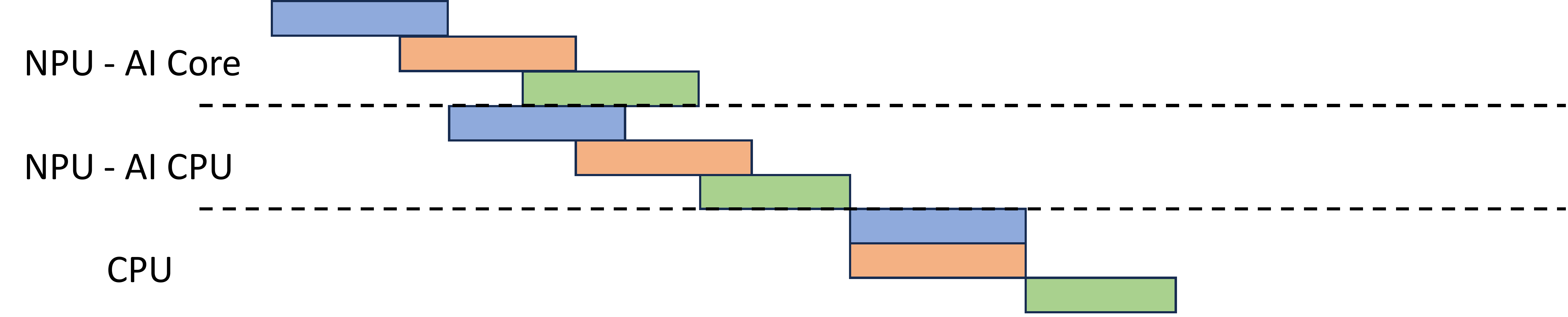}
        \label{fig:pipeline_npu}
    }\\[5pt]
    \subfloat[Cross-device collaborative execution across AI Core, AI CPU and host CPU]{
        \includegraphics[width=0.95\columnwidth, height=3.5cm, keepaspectratio]{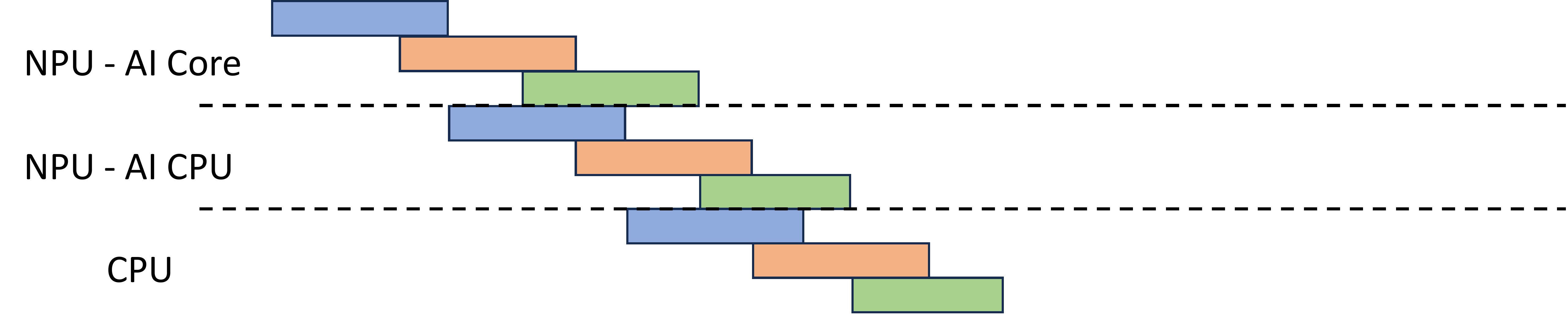}
        \label{fig:pipeline_npu_cpu}
    }
    \caption{NPU-CPU collaboration design across different hardware levels. (a) Baseline design with no pipeline parallelism, each stage executes sequentially leading to significant latency overhead. (b) Optimized design with pipeline parallelism between NPU internal AI Core and AI CPU, masking Top-$k$ selection latency. (c) Cross-device collaborative execution, overlapping distance calculation on AI Core, Top-$k$ selection on AI CPU, and re-ranking on host CPU to maximize end-to-end throughput under high load.}
    \label{fig:pipeline_overview}
\end{figure}

As shown in Figure \ref{fig:pipeline_overview} (c), this collaborative execution path overlaps distance calculation on AI Core, Top-$k$ selection on AI CPU, and final re-ranking on host CPU within each query batch. We expect this optimization to improve end-to-end throughput significantly under high load once the engineering implementation is complete.

\subsection{Distributed Scaling}
As vector dataset sizes continue to grow or throughput requirements increase further, single NPU devices will 
eventually encounter bottlenecks in memory capacity or computing resources. 
To address this challenge, Ascend-RaBitQ is designed with native distributed scaling capability. 
We adopt a cluster-aware load balancing strategy to evenly partition quantized vector datasets across 
multiple NPU devices, ensuring that the number of vectors in each inverted list is balanced across all nodes. 
During the index construction phase, each NPU independently performs quantization operations on its responsible vector 
shard in parallel, allowing construction throughput to scale with the number of devices. During search, 
multiple NPU devices simultaneously process the same query batch, independently performing their local coarse 
search computation tasks. Finally, the CPU aggregates results from all NPU nodes and performs global re-ranking to output 
the final accurate Top-$k$ search results.

\section{Evaluation}
\label{sec:evaluation}

We conduct extensive experiments to evaluate the performance of our proposed Ascend-RaBitQ system on a variety of standard 
vector search datasets, comparing against state-of-the-art CPU IVF quantization implementations and the native NPU 
full-precision baseline. All experiments are performed on a server equipped with 8 Ascend NPUs and a host CPU. Each NPU features 48 AIV cores and 24 AIC cores clocked at 1.85~GHz, with 64~GB HBM. The host CPU has 192 cores at 2.6~GHz and 1.5~TB DDR4 memory. Unless otherwise specified, experiments use a single NPU together with the host CPU.

\subsection{Experimental Setup}

Datasets. We use four widely adopted public datasets covering different application domains (image, text) and data 
scales ranging from 1 million to 1 billion vectors:
\begin{itemize}
    \item SIFT1M~\cite{jegou2011product}: 1 million 128-dimensional SIFT descriptors for image retrieval (small-scale image workload);
    \item Cohere 10M~\cite{cohere2023vectordbbench10m}: 10 million 768-dimensional text embeddings (large-scale text semantic search workload);
    \item SIFT100M~\cite{jegou2011product}: 100 million 128-dimensional SIFT descriptors (medium-scale hundred-million-level candidate image workload);
    \item SIFT1B~\cite{johnson2019billion}: 1 billion 128-dimensional SIFT descriptors (large-scale billion-level production workload).
\end{itemize}

Baselines. We compare against two categories of state-of-the-art baselines for fair evaluation:
\begin{enumerate}
    \item Same NPU platform baseline: Ascend IndexSDK IVF-Flat implementation, the native full-precision vector search solution 
    on Ascend NPU;
    \item CPU platform baselines: Three standard IVF-RaBitQ implementations from FAISS library~(v1.14.1)~\cite{douze2024faiss} with different quantization configurations, each optimized with architecture-specific instructions for the target CPU platform:
    \begin{itemize}
        \item \textbf{CPU 1-bit}: Adopts 1-bit RaBitQ quantization for database vectors only, keeping query vectors in full precision. This variant is \textit{mathematically equivalent} to our Ascend-RaBitQ, making it the fairest baseline for evaluating the raw compute advantage of NPU acceleration over CPU.
        \item \textbf{CPU 1-bit+SQ}: Applies 4-bit scalar quantization (SQ) to query vectors while keeping database vectors at 1-bit RaBitQ quantization. The query-side SQ enables the CPU to efficiently compute inner products between two low-precision vectors using SIMD instructions, significantly accelerating distance computation.
        \item \textbf{CPU 4-bit+SQ}: Quantizes database vectors to 4-bit using RaBitQ and applies 4-bit scalar quantization (SQ) to query vectors. This configuration achieves higher precision than the 1-bit variants at the cost of increased memory access and computation, serving as a reference for balanced quantization-performance tradeoffs.
    \end{itemize}
\end{enumerate}

All methods adopt identical IVF configurations for fair comparison, with $nList = \sqrt{N}$ and $nProbe = 5\% \cdot nList$, 
where $N$ denotes the total number of vectors in the dataset. 
We adopt recall@10, recall@100, recall@300, queries per second (QPS), index construction time, and memory footprint as the primary evaluation metrics.
To facilitate intuitive cross-method comparison within each figure, all QPS values are locally normalized: the minimum QPS value within each figure is set to 1, and all other values are reported as relative results against this baseline.
This setup separates platform-native comparison against full-precision NPU IVF-Flat from algorithm-equivalent comparison against CPU IVF-RaBitQ. The former quantifies the benefit of replacing full-precision NPU search with compact RaBitQ search on the same platform, while the latter isolates the benefit of mapping the same 1-bit estimator onto NPU execution units.

\subsection{NPU Platform Native Comparison: Ascend-RaBitQ vs. IVF-Flat}
We conduct a comprehensive performance comparison between Ascend-RaBitQ and the native Ascend IndexSDK IVF-Flat implementation 
on the Ascend NPU platform. An important practical consideration motivates our experimental design here: 
the full-precision IVF-Flat index requires storing the entire dataset in NPU memory, which imposes significant 
memory constraints. For a single Ascend NPU with 64~GB HBM, only the SIFT1M dataset (489~MB full-precision index) 
can be accommodated. Larger datasets such as SIFT100M (48~GB) and SIFT1B (480~GB) exceed single-NPU memory capacity, making single-NPU IVF-Flat comparisons infeasible. Therefore, we first compare Ascend-RaBitQ against 
IVF-Flat on a single NPU using the SIFT1M dataset, then extend the comparison to four NPUs using the SIFT100M dataset.

\begin{figure}[htbp]
    \centering
    \includegraphics[width=0.95\linewidth]{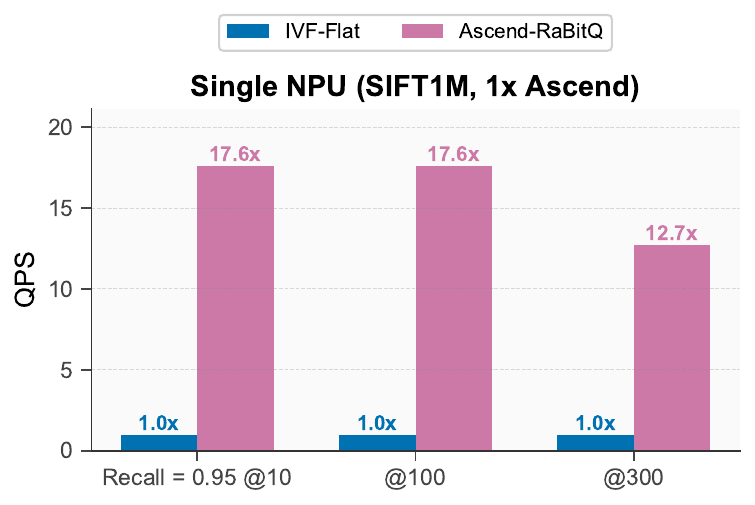}
    \caption{Single-NPU performance comparison on SIFT1M dataset (1$\times$ Ascend).}
    \label{fig:npu_single_card}
\end{figure}

Figure \ref{fig:npu_single_card} presents the single-NPU (1$\times$ Ascend) performance comparison on the SIFT1M dataset. 
Ascend-RaBitQ demonstrates consistent performance improvements across all three Top-$k$ settings at the same recall. 
Ascend-RaBitQ achieves a 17.6$\times$ QPS speedup over IVF-Flat at recall = 0.95 under both top-10 and top-100 settings, 
and 12.7$\times$ under top-300. This performance advantage originates from two key architectural factors. 
First, our 1-bit quantization reduces memory access volume by 32$\times$ compared to full-precision IVF-Flat, effectively 
eliminating the HBM bandwidth bottleneck even when retrieving significantly more candidates for higher Top-$k$ targets. 
Second, while both methods leverage the Cube Unit for matrix multiplication-based distance computation, Ascend-RaBitQ's 
1-bit quantized vectors enable substantially lower computational complexity than IVF-Flat's full-precision 
vector distance calculation. 

For real-time query latency on a single NPU, Ascend-RaBitQ achieves an average latency of 0.34~ms with a small batch size of 64, 
representing an order-of-magnitude reduction compared to IVF-Flat's 5.98~ms latency. This enables the system to meet strict low-latency 
online serving requirements. 
In terms of memory footprint, Ascend-RaBitQ achieves a 32$\times$ theoretical compression ratio over \texttt{FP32}. For the SIFT1M dataset, the IVF-Flat index occupies 489~MB and must reside entirely in NPU memory during query processing, while the Ascend-RaBitQ index requires only 20~MB including 1-bit index codes and precomputed values, achieving a 95.9\% reduction in memory usage. 

\begin{figure}[htbp]
    \centering
    \includegraphics[width=0.95\linewidth]{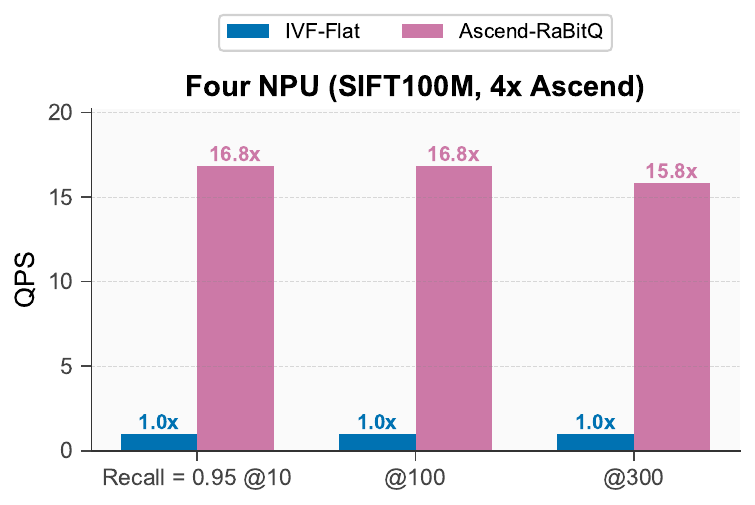}
    \caption{Four-NPU performance comparison on SIFT100M dataset (4$\times$ Ascend).}
    \label{fig:npu_four_card}
\end{figure}

To evaluate performance on larger datasets, we extend our comparison to a four-NPU configuration using the SIFT100M dataset. 
With four Ascend NPUs providing an aggregated 256~GB HBM capacity, the 48~GB SIFT100M full-precision index can be 
fully accommodated through data parallel partitioning across devices. Figure \ref{fig:npu_four_card} presents the 
four-NPU performance comparison results. 
At recall = 0.95, Ascend-RaBitQ achieves a 16.8$\times$ QPS speedup over IVF-Flat for both top-10 and top-100, 
and 15.8$\times$ for top-300. The slightly higher speedup on SIFT100M compared to 
SIFT1M can be attributed to improved computational efficiency at larger scales, where the quantization benefits become 
more pronounced as memory bandwidth constraints intensify. Furthermore, the memory compression advantage of Ascend-RaBitQ 
becomes even more critical for billion-scale datasets: while SIFT1B would require 480~GB of memory for IVF-Flat, 
exceeding even the 256~GB aggregated capacity of four NPUs, Ascend-RaBitQ requires only 20~GB, comfortably fitting within 
a single NPU and eliminating expensive inter-device communication overhead.

\subsection{Cross-platform Comparison: Ascend-RaBitQ vs. CPU IVF-RaBitQ Implementations}
To demonstrate the cross-platform performance advantages of our NPU-accelerated approach, we compare Ascend-RaBitQ against three standard CPU IVF-RaBitQ implementations from the FAISS library under the same IVF configuration but different quantization choices: 1-bit, 1-bit + query SQ, and 4-bit + query SQ. This comparison uses CPU 1-bit as the algorithm-equivalent baseline and CPU 1-bit+SQ as the strongest CPU throughput baseline, since query SQ makes CPU SIMD execution substantially more efficient.

\begin{figure*}[htbp]
    \centering
    \includegraphics[width=\linewidth]{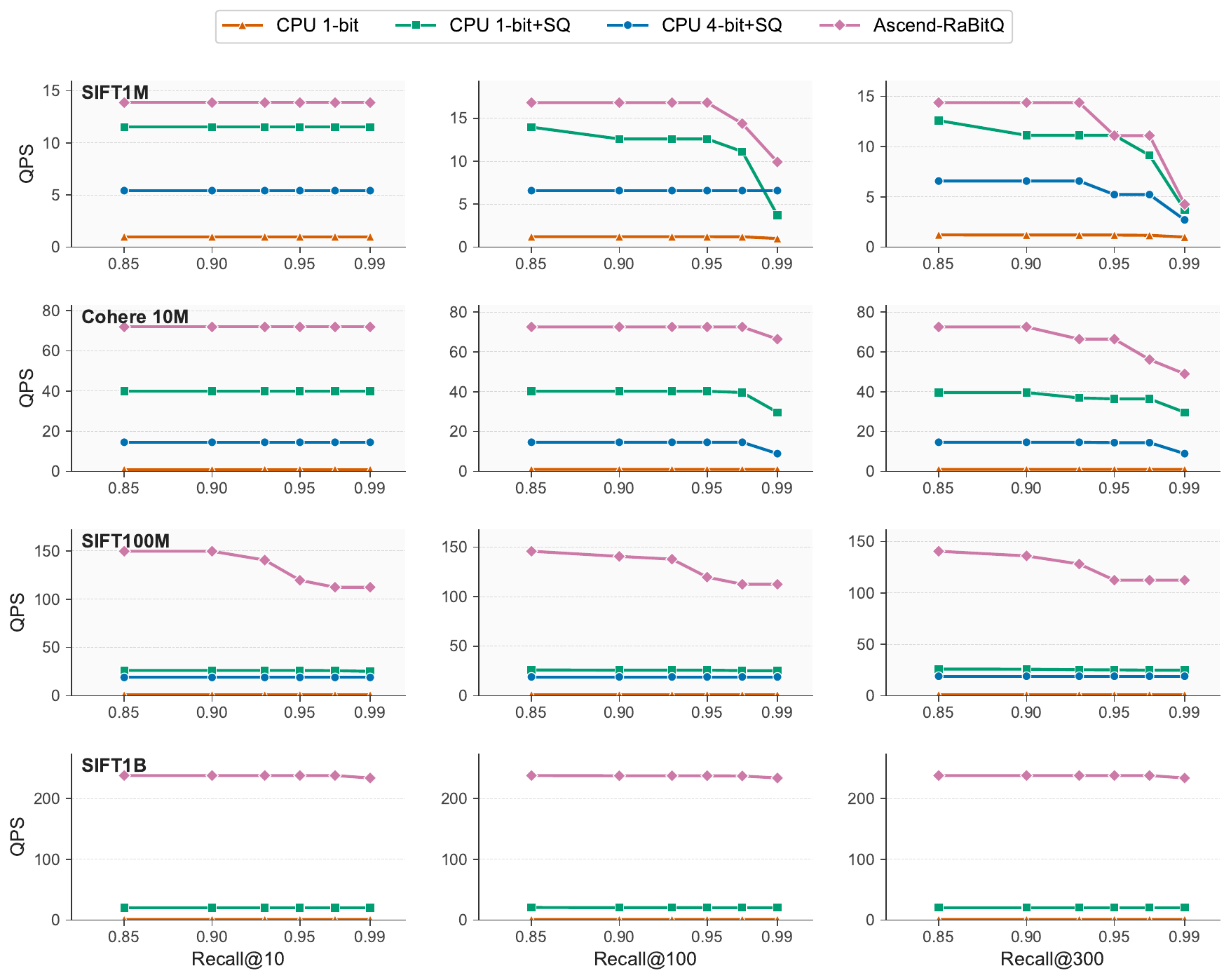}
    \caption{Cross-platform performance comparison across different recall thresholds. Each column corresponds to a recall condition (Recall@10, Recall@100, Recall@300), and each row corresponds to a dataset. Curves show QPS versus recall for each method under different recall conditions.}
    \label{fig:cpu_performance_compare}
\end{figure*}

Figure \ref{fig:cpu_performance_compare} presents the QPS comparison across four datasets and three recall 
conditions (Recall@10, Recall@100, Recall@300). The four rows correspond to the four datasets, 
while the three columns correspond to the three recall conditions. Each line chart shows the QPS-recall 
tradeoff for all available methods under a given recall condition, with lines connecting the data points to 
illustrate the performance trend as recall varies. Ascend-RaBitQ demonstrates consistent performance advantages 
across all CPU configurations and datasets, with the speedup over the fastest CPU implementation growing with 
dataset scale. On the small-scale SIFT1M dataset, Ascend-RaBitQ achieves a 1.2$\times$ speedup over the best 
CPU implementation (CPU 1-bit+SQ) at Recall@10, though the gap narrows at higher Top-$k$ settings: at Top-300, 
CPU 1-bit+SQ nearly matches Ascend-RaBitQ, indicating that query-side scalar quantization can partially 
compensate when retrieving a very large number of candidates. On the larger SIFT100M dataset, the advantage becomes 
more pronounced, with Ascend-RaBitQ delivering a 4.6$\times$ speedup over CPU 1-bit+SQ. This trend confirms that the 
Ascend NPU's computation unit provides greater relative efficiency for vector search as memory bandwidth 
pressure intensifies with larger datasets. On the high-dimensional Cohere 10M dataset (768-dimensional text embeddings), 
Ascend-RaBitQ maintains a stable 1.8$\times$ speedup over CPU 1-bit+SQ across all Top-$k$ settings, demonstrating 
robust performance on text-based semantic search workloads. On the billion-scale SIFT1B dataset, 
Ascend-RaBitQ achieves approximately 11.7$\times$ higher throughput than CPU 1-bit+SQ and over 238$\times$ 
over CPU 1-bit, validating its practicality for billion-scale deployments. Note that we exclude CPU 4-bit+SQ from the SIFT1B evaluation as its index construction failed to complete within one week.

Across all configurations, the CPU 1-bit variant without query-side quantization trails by one to two orders of magnitude: 
Ascend-RaBitQ outperforms it by 13.9$\times$ on SIFT1M, by over 100$\times$ on SIFT100M, and by over 238$\times$ on SIFT1B. Critically, this 
performance gap does not stem from memory bandwidth limitations but from the absence of query SQ, which prevents 
CPU from efficiently computing inner products between two 1-bit quantized vectors using SIMD instructions. 
It is worth emphasizing that our Ascend-RaBitQ also does not apply query-side SQ and therefore remains mathematically 
equivalent to the CPU 1-bit variant, offering higher precision than methods that incorporate query SQ. 
The stark performance advantage over the mathematically equivalent CPU 1-bit implementation thus reflects the fundamentally 
superior compute throughput of the NPU's execution units. 
Extending query-side SQ optimization to the NPU pipeline represents a promising direction for future work to further improve 
end-to-end throughput.

Beyond query performance, we further compare the end-to-end index construction time across all datasets, 
as shown in Table \ref{tab:index_time}. Ascend-RaBitQ achieves a speedup of 3.0$\times$ to 62.8$\times$ over the 
CPU 1-bit IVF-RaBitQ implementation. This improvement comes from offloading both the $K$-Means 
clustering step for IVF partition and the 1-bit quantization encoding step to the Ascend NPU's execution units. The result is important for production deployment because index construction is not a one-time cost: vector databases rebuild or refresh IVF partitions as corpora grow, embeddings are updated, or hot subsets are re-indexed.

\begin{table}[!htbp]
    \centering
    \resizebox{\linewidth}{!}{
    \begin{tabular}{lccc}
    \toprule
    Dataset & CPU 1-bit IVF-RaBitQ & Ascend-RaBitQ & Speedup \\
    \midrule
    SIFT1M & 1.65 min & 6.87 s & 14.42$\times$ \\
    Cohere 10M & 1.55 h & 1.48 min & 62.84$\times$ \\
    SIFT100M & 4.64 h & 1.53 h & 3.03$\times$ \\
    SIFT1B & 98.63 h & 4.20 h & 23.49$\times$ \\
    \bottomrule
    \end{tabular}
    }
    \caption{Index construction time comparison across all datasets.}
    \label{tab:index_time}
\end{table}

\subsection{Ablation Study}
\begin{figure}[htbp]
    \centering
    \includegraphics[width=0.95\linewidth]{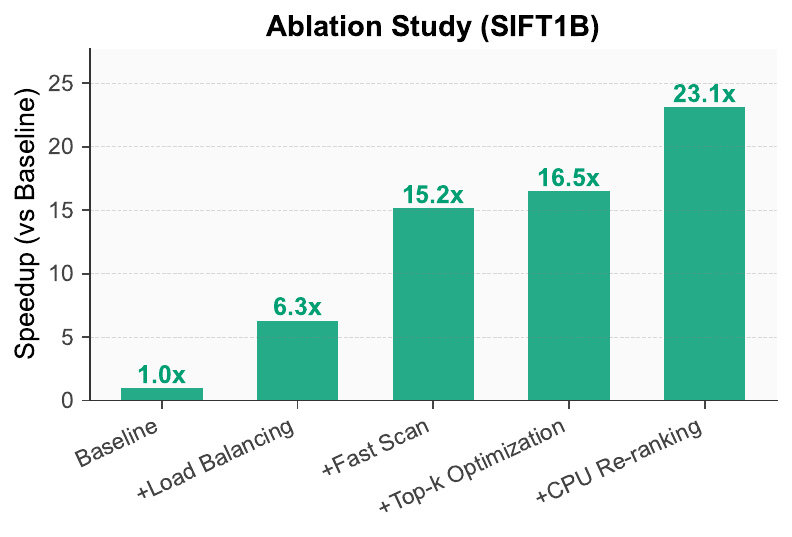}
    \caption{Ablation study results on SIFT1B dataset, showing speedup of different optimization combinations vs. the naive baseline.}
    \label{fig:ablation_study}
\end{figure}

We validate the Section~3.3 search design through controlled ablations on SIFT1B, as shown in Figure \ref{fig:ablation_study}. The naive baseline uses CANN \texttt{select} for distance computation, schedules work at query granularity, executes distance calculation and Top-$k$ sequentially, and performs re-ranking on the NPU. This setup exposes the four bottlenecks targeted by our design: AIV under-utilization, inefficient 1-bit arithmetic, Top-$k$ latency, and raw-vector data movement. We then add each optimization incrementally. Index block-level load balancing (Section~3.3.1) removes idle AIV cores caused by uneven query-level launches and reaches a 6.3$\times$ speedup. Replacing \texttt{select} with gather-based FastScan (Section~3.3.2) uses precomputed LUTs for compact-code distance estimation and raises the cumulative speedup to 15.2$\times$. AI CPU Top-$k$ pruning with AI Core-AI CPU pipeline parallelism (Section~3.3.3) masks nearly all selection latency and reaches 16.5$\times$. Finally, CPU re-ranking (Section~3.3.4) avoids gathering discontinuous raw vectors into NPU-contiguous buffers, yielding the final 23.1$\times$ speedup.

Overall, the ablation results show that Ascend-RaBitQ's performance comes from hardware-matched design choices rather than a single kernel optimization: compact-code distance estimation is converted from element-wise masking to LUT-based gather, and fine re-ranking is placed where raw vectors already reside. This supports our central claim that NPU acceleration requires joint redesign of data layout, operator mapping, scheduling, and CPU-NPU stage placement.

\subsection{Multi-NPU Large-scale Scalability Test}
\begin{figure}[htbp]
    \centering
    \includegraphics[width=0.95\linewidth]{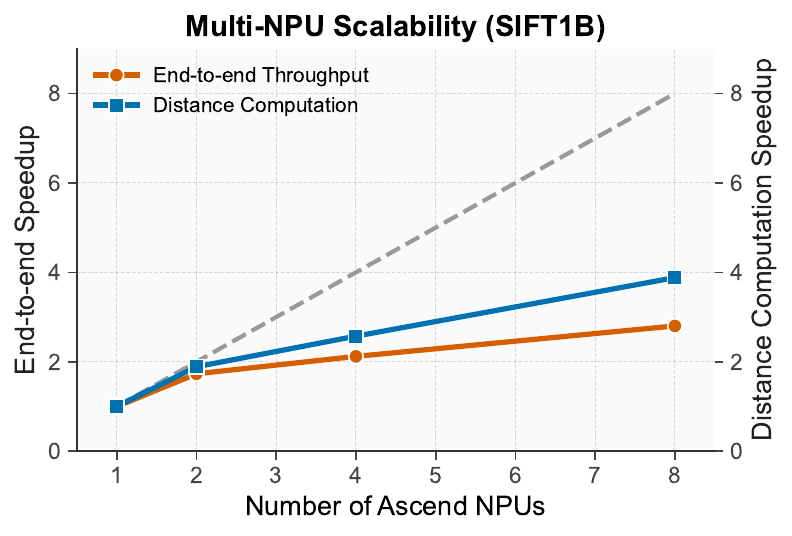}
    \caption{Scalability of Ascend-RaBitQ with increasing number of Ascend NPUs on SIFT1B dataset. The left axis shows end-to-end throughput speedup, while the right axis shows raw distance computation throughput speedup. The gap between NPU-side distance-computation scaling and end-to-end scaling indicates a bottleneck shift to CPU post-processing stages.}
    \label{fig:multicard_scalability}
\end{figure}

Finally, we evaluate the scalability of Ascend-RaBitQ on the billion-scale SIFT1B dataset using up to 8 Ascend NPUs to demonstrate its suitability for large-scale industrial deployment. Figure \ref{fig:multicard_scalability} presents the speedup of end-to-end throughput and raw distance computation throughput as the number of NPUs increases. As shown in the figure, the end-to-end throughput achieves approximately 3$\times$ speedup on 8 NPUs compared to a single NPU, falling short of linear scaling. This is because NPU acceleration primarily benefits the coarse ranking stage, while the overall throughput remains constrained by CPU re-ranking. As more NPUs are added, the system bottleneck progressively shifts from NPU computation to CPU post-processing. This finding suggests that tighter NPU-CPU collaboration that overlaps re-ranking with NPU computation could further improve end-to-end throughput. The raw distance computation stage scales better than end-to-end throughput, achieving approximately 4$\times$ speedup on 8 NPUs, but it also remains sub-linear due to the scheduling and load-balance effects analyzed below. Thus, the two curves should be interpreted separately: one reflects system-level bottleneck migration, while the other isolates NPU-side scalability.

Our in-depth analysis of the experimental data reveals two primary causes for the sub-linear scaling of distance computation. First, we manage vectors at the granularity of index blocks, where the total number of blocks determines the number of kernel launches. Even when inverted lists are distributed across different NPUs, the total number of blocks may not decrease linearly, leading to sub-optimal kernel launch efficiency at scale. Second, the vector counts across inverted lists are inherently uneven due to the data distribution. Although we employ block-level load balancing, when blocks from different lists are computed concurrently, inter-core load imbalance can still occur, further reducing linearity.

Two potential directions can address these issues. First, introducing list-level load balancing during clustering training could ensure a more balanced vector distribution across lists. Second, using a smaller block size would improve the granularity of block-level load balancing. However, smaller blocks lead to more fragmented memory storage, degrading memory access efficiency, and the increased number of blocks adds management complexity. Determining the optimal block size for different workload scenarios is an interesting research problem and remains as our future work.

\section{Related Work}
\label{sec:related_work}

We review related work in two categories: large-scale ANNS systems and hardware acceleration for ANNS.

\subsection{Large-scale ANNS Systems}

Billion-scale ANNS systems have evolved along three directions: SSD-based approaches to overcome memory constraints, 
heterogeneous architectures to accelerate computation, and distributed deployment to scale storage and throughput.

\textbf{SSD-based systems.} DiskANN~\cite{jayaram2019diskann} stores graph indexes on SSDs, serving billion-scale queries 
with low latency on a single workstation. FlashANNS~\cite{xiao2025breaking} proposes a GPU-driven asynchronous I/O framework 
that overlaps SSD reads with GPU computation, significantly improving throughput over DiskANN. 
Its insight---pipelining I/O and computation through dependency relaxation---is orthogonal to our work: we maximize on-chip 
compute utilization once data resides in HBM.

\textbf{Heterogeneous systems.} FusionANNS~\cite{tian2025towards} proposes CPU/GPU collaborative filtering and 
re-ranking using SSDs and a single entry-level GPU, achieving substantial throughput gains over SSD-based baselines. 
GustANN~\cite{jiang2025high} adopts a GPU-centric, CPU-assisted architecture with memory-efficient kernels, 
outperforming prior SSD-based systems.

\textbf{Distributed systems.} SOGAIC~\cite{shi2025scalable} introduces overload-aware partitioning and 
load-balanced distributed sub-graph construction, substantially reducing index build time for 10-billion-scale corpora. 
DISTRIBUTEDANN~\cite{adams2025distributedann} scales a DiskANN graph across thousands of machines to store and search tens of 
billions of vectors with low latency at high throughput, now powering Bing.

These systems share a philosophy with Ascend-RaBitQ: no single hardware platform optimally serves all pipeline stages. 
However, they target CPU-GPU-SSD hierarchies and do not address NPU-specific characteristics---heterogeneous compute units, 
software-managed on-chip buffers, and asynchronous data movement engines. 
Ascend-RaBitQ addresses this gap by exploiting these features for billion-scale vector search, achieving high hardware utilization 
through algorithm-hardware co-optimization.

\subsection{Hardware Acceleration for ANNS}

Substantial effort has optimized individual hardware platforms for ANNS---from CPUs and GPUs to the emerging NPUs targeted 
by our work.

\textbf{CPU acceleration.} FAISS~\cite{douze2024faiss} has been widely deployed on CPUs. KBest~\cite{ma2026kbest} tailors 
vector search for the ARM-based Kunpeng 920, incorporating SIMD acceleration, data prefetch, and early termination to 
achieve significant throughput gains over x86-optimized libraries. At the algorithm level, RaBitQ supports flexible compression 
rates~\cite{gao2025practical} with asymptotic optimality guarantees. 
However, even highly tuned CPU implementations remain constrained by memory bandwidth and compute throughput at billion scale.

\textbf{GPU acceleration.} Johnson et al.~\cite{johnson2019billion} established the foundational GPU-accelerated framework. 
IVF-RaBitQ (GPU)~\cite{shi2026ivfrabitq} integrates RaBitQ into a GPU-native IVF pipeline with fused search kernels, 
outperforming graph-based GPU methods at equivalent recall. Tagore~\cite{li2025scalable} accelerates graph index construction 
via GNN-Descent and an asynchronous GPU-CPU-disk framework, achieving substantial speedup over CPU construction.

\textbf{CPU-GPU cooperation.} PilotANN~\cite{gui2026pilotann} decomposes graph search into GPU subgraph traversal and CPU 
refinement, handling datasets far larger than GPU memory. SVFusion~\cite{peng2026svfusion} employs a GPU-CPU-disk framework 
with hierarchical indexing, achieving high throughput with online updates. RUMMY~\cite{zhang2024fast} pipelines data 
transmission and GPU computation via query-aware reordering, significantly outperforming IVF-GPU.

\textbf{NPU acceleration.} ANNS on NPUs remains largely underexplored. GaussDB-Vector~\cite{sun2025gaussdb} demonstrates a 
large-scale vector database for LLM applications, but does not specifically optimize for NPU architectures. 
To the best of our knowledge, no prior work has realized RaBitQ on commercial NPUs or addressed the full path of billion-scale IVF-RaBitQ search, including compact-code distance computation, Top-$k$ selection, and CPU-NPU re-ranking placement.

Our work fills this gap by implementing and optimizing RaBitQ on commercial NPUs, leveraging 
fused AIC-AIV operators, computation flow restructuring, compact-code lookup design, resource scheduling, and NPU-CPU 
collaboration to achieve high hardware utilization and throughput.

\section{Conclusion}
\label{sec:conclusion}

This paper presents Ascend-RaBitQ, to our knowledge, the first IVF-RaBitQ system optimized for commercial NPUs. By decomposing vector search into a three-stage collaborative execution path---AI Core coarse ranking, AI CPU Top-$k$ selection, and host CPU fine re-ranking---our approach maps each stage to its optimal hardware, resolving the accuracy-memory-performance trilemma. Extensive experiments demonstrate that Ascend-RaBitQ achieves 3.0$\times$ to 62.8$\times$ faster index construction, up to 11.7$\times$ throughput improvement over the fastest CPU implementation, and over two orders of magnitude over the mathematically equivalent CPU baseline. Controlled ablation studies confirm a cumulative 23.1$\times$ speedup from the implemented optimizations, and multi-NPU tests demonstrate scalable throughput on distributed configurations.

We are actively extending Ascend-RaBitQ to trillion-scale cross-NPU-node deployments, deepening integration with production vector databases such as Milvus~\cite{2021milvus} and Faiss~\cite{douze2024faiss}, and optimizing for low-latency single-query scenarios.

\bibliographystyle{ACM-Reference-Format}
\bibliography{tex/references}

\end{document}